\documentclass[10pt,journal]{IEEEtranTCOM}

\usepackage[english]{babel}
\usepackage[usenames]{color}
\usepackage[cp1250]{inputenc}
\usepackage{amsfonts}
\usepackage{amsthm}
\usepackage{graphicx}
\usepackage{epsfig}
\usepackage{mathrsfs}
\usepackage{amsmath}
\usepackage{algorithm}
\usepackage{algorithmic}
\usepackage{hyperref}
\usepackage{mathtools}

\theoremstyle{plain}

\begin{document}
\newcommand{\bea}{\begin{eqnarray}}
\newcommand{\eea}{\end{eqnarray}}
\newcommand{\be}{\begin{equation}}
\newcommand{\ee}{\end{equation}}
\newcommand{\beas}{\begin{eqnarray*}}
\newcommand{\eeas}{\end{eqnarray*}}
\newcommand{\bs}{\backslash}
\newcommand{\bc}{\begin{center}}
\newcommand{\ec}{\end{center}}
\def\SC {\mathscr{C}}

\title{{3-SAT solver for two-way quantum computers}}
\author{\IEEEauthorblockN{Jarek Duda}\\
\IEEEauthorblockA{Faculty of Mathematics and Computer Science, Jagiellonian University, Krakow, Poland, \emph{dudajar@gmail.com}}}
\maketitle

\begin{abstract}
While quantum computers assume existence of \emph{state preparation} process $|0\rangle$, CPT symmetry of physics says that performing such process in CPT symmetry perspective, e.g. reversing used EM impulses ($V(t)\to V(-t)$), we should get its symmetric analog $\langle 0|$, referred here as \emph{state postparation} - which should provide results as \emph{postselection}, but with higher success rate.

Two-way quantum computers (2WQC) assume having both $|0\rangle$ and $\langle 0|$ pre and postparation. In theory they allow to solve NP problems, however, basic approach would be more difficult than Shor algorithm, which is now far from being practical. This article discusses approach to make practical 2WQC 3-SAT solver, requiring exponential reduction of error rate, what should be achievable through linear increase of the numbers of gates. 

2WQC also provides additional error correction capabilities, like more stable Grover algorithm, or mid-circuit enforcement of syndrome to zero, like proposed equalizer enforcing qubit equality.
\end{abstract}
\section{Introduction}
While CPT symmetry is necessary for Lorentz invariant local quantum field theories~\cite{CPT}, standard one-way quantum computers (1WQC) assume asymmetric treatment of boundary conditions: allowing only state preparation at the beginning, and only measurement at the end.

For QC technologies like silicon quantum dots~\cite{dots}, state preparation $|0\rangle$ is performed with EM impulses - hence using reversed impulses ($V(t) \to V(-t)$), we should perform state preparation in perspective of CPT symmetry, what corresponds to $\langle 0|$. While pre-measurement is considered in literature~\cite{premeasure}, alone it does not seem to bring new capabilities. In contrast, adding state postparation $\langle 0|$ as reversed preparation brings essential improvements, like 3-SAT solver we will discuss here. 

Therefore, we will focus on 2WQC as 1WQC with added $\langle 0|$ state postparation operation, which mathematically gives results as postselection $|0\rangle\langle 0|$, but with higher success rate - physically enforcing the final state exactly as state preparation enforces the initial state. For example 2WQC allows for faster and more stable Grover algorithm~\cite{Grover}, and generally should allow to solve PostBQP~\cite{bqp} problems. However, while cloning is claimed being possible using closed timelike curves~\cite{CTC}, for 2WQC no-cloning theorem remains valid~\cite{noclon}. 

\begin{figure}[t!]
    \centering
        \includegraphics[width=9cm]{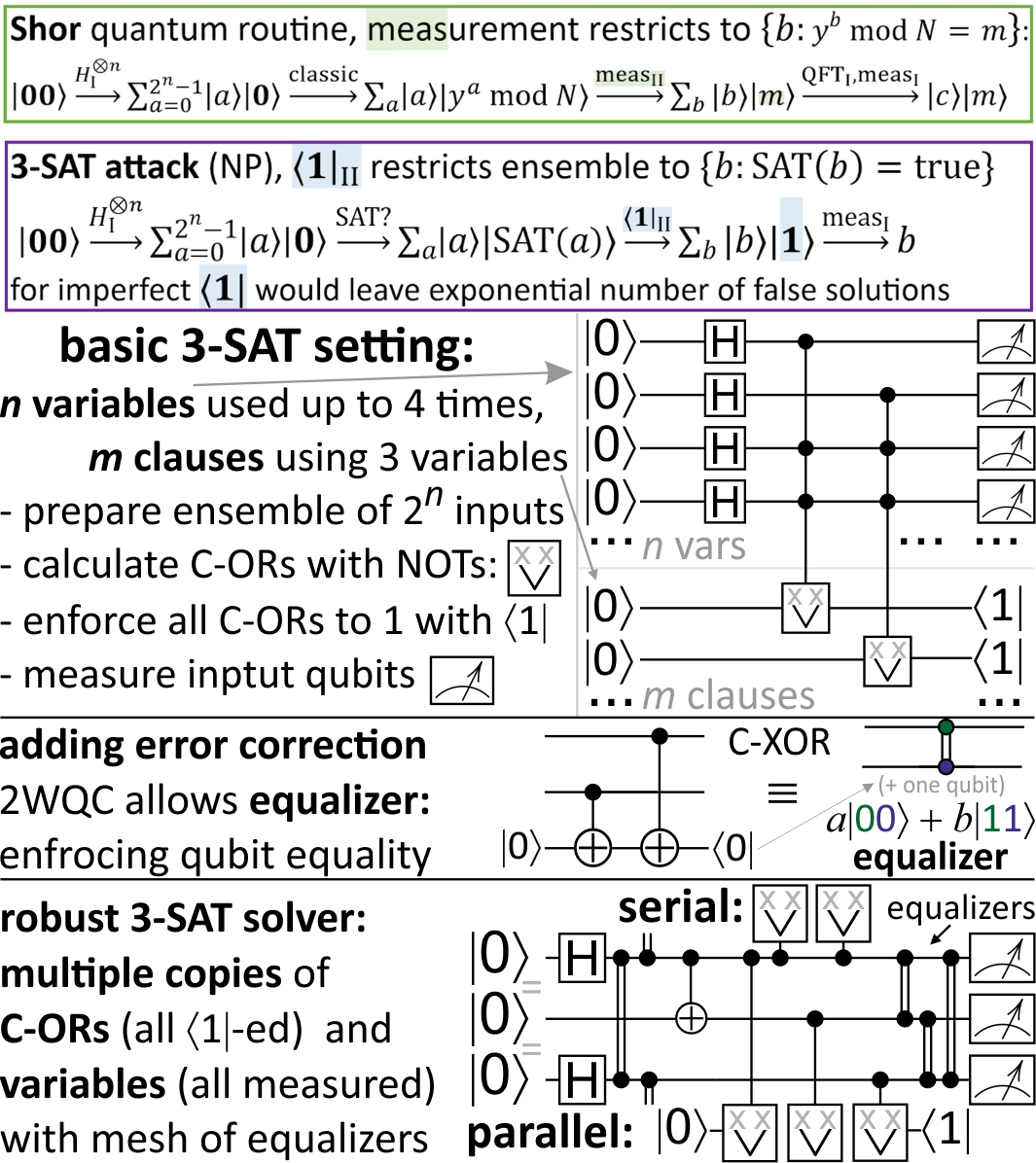}
        \caption{Top: Shor algorithm prepares superposition of $2^n$ inputs, calculates classical function for them and measure value, also measure Fourier transform of input. Proposed 3-SAT solver replaces this value measurement with $\langle 1|$ postparation as CPT analog of state preparation, allowed as additional operation in 2WQC. Center: basic scheme of such approach, using C-ORs as 4-qubit gates: $(x,y,z,t)\to (x,y,z,t \oplus ((\neg)x \vee (\neg)y \vee (\neg)z))$, with negated some variables. Bottom: adding error correction mechanisms, e.g. using the shown equalizer gadget, and finally for 3-SAT there are rather required multiple copies of C-ORs: in serial setting, or parallel to lower the depth - where it seems useful to apply mesh of equalizers to maintain equality of multiple copies of variables. }
        \label{3SAT}
\end{figure}

As summarized in Fig. \ref{3SAT}, this article focuses on 3-SAT solver initially suggested in \cite{2WQC}. Generally it seems more difficult to realize than Shor algorithm~\cite{shor}, currently being far from practical. However, 2WQC also brings looking more powerful error correction capabilities, like proposed equalizer gadget, as discussed here in theory allowing to overcome imperfections, and hopefully in future leading to practical realizations.

\section{3-SAT solver for two-way quantum computers}
\begin{figure}[t!]
    \centering
        \includegraphics{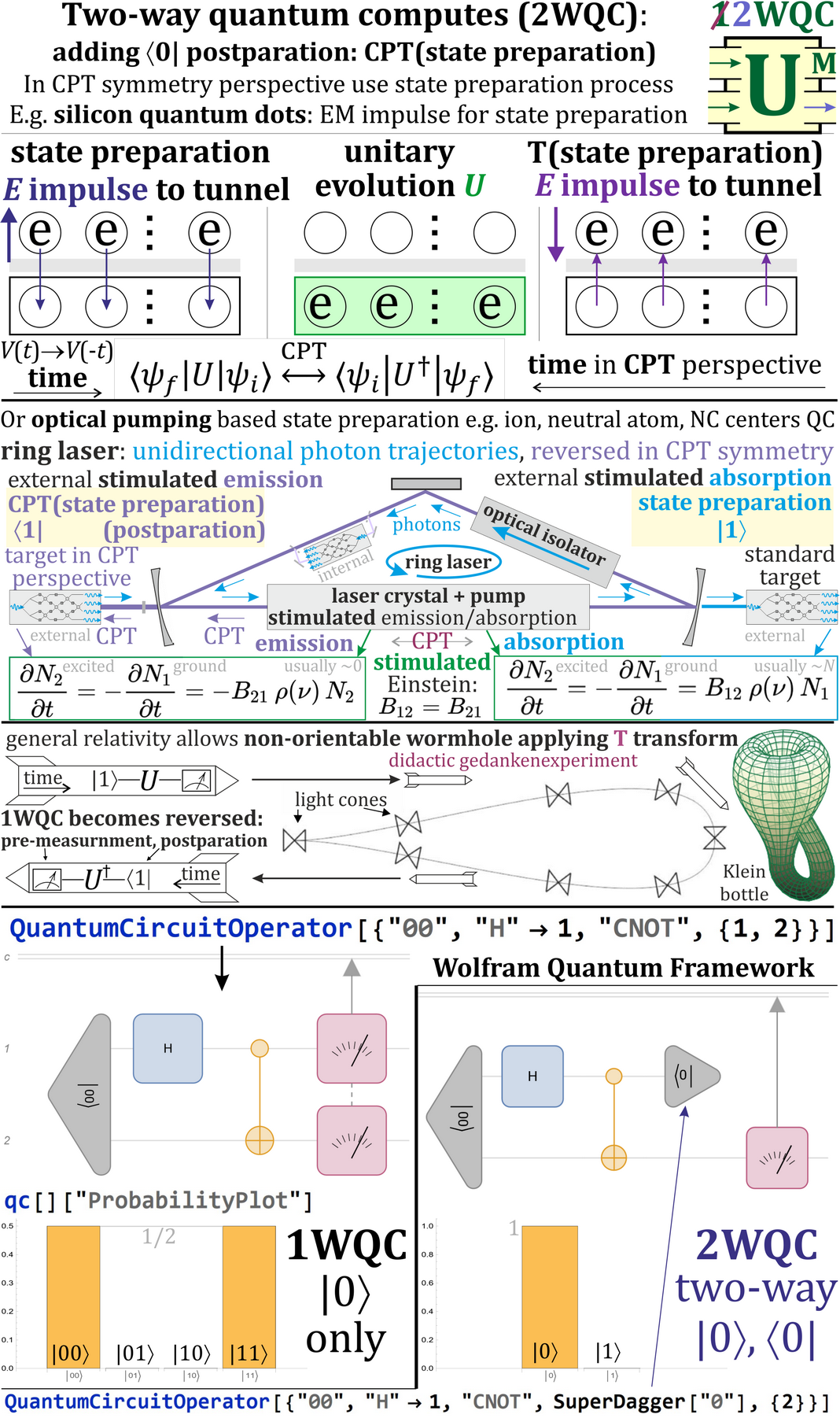}
        \caption{2WQC~\cite{2WQC} proposes to extend 1WQC set of allowed operation by CPT analog of state preparation (like postselection but by physical constraints) - performing state preparation process in CPT perspective, e.g. by reversing used EM impulses (top), or using ring laser with optical isolator (center). There is also shown didactic gedankenexperiment with allowed by general relativity Klein-bottle like wormhole~\cite{Klein} applying T symmetry: 1WQC in a rocket travelling through it would be reversed: start with postparation, end with pre-measurement. Bottom: simple example of implementation in Wolfram Quantum Framework. }
        \label{2WQC}
\end{figure}

This main Section introduce to 2WQC (summarized in Fig. \ref{2WQC}), then discuss proposed 3-SAT solver and adding error correction.
\subsection{Postselection $|0\rangle \langle 0|$ and conjugated state preparation $\langle 0|$} 
For density matrix $\rho$, e.g. near the end of quantum computation, standard 1WQC measurement has probability of state $s$:
\be \textrm{1WQC:}\qquad P(s)=\textrm{Tr}(\Pi_s\,\rho) \ee
for $\Pi_s$ projection, e.g. $\Pi_s=\otimes_{i=1..n} |s_i\rangle \langle s_i|$ for $n$ qubits.

In p1WQC as postselected 1WQC to condition $c$, e.g. choosing values of some additional $m$ qubits ($\rho$ increased from $n$ to $n+m$ qubits), we still use the above formula, but discard results not satisfying this condition. After normalization (we have to assume $P(c)>0$) we analogously get (e.g. $I_s=I_{2^n}$):
\be \textrm{p1WQC:}\qquad P(s|c)=\frac{\textrm{Tr}((\Pi_s\otimes \Pi_c)\,\rho)}{P(c)} \ee
$$ \textrm{for}\quad P(c) = \sum_s \textrm{Tr}((\Pi_s\otimes \Pi_c)\,\rho) = \textrm{Tr}((I_s\otimes \Pi_c)\,\rho) $$
being probability of condition $c$, which is $\sim 2^{-m}$ assuming uniform distribution, and e.g. $\Pi_c=\otimes_{i=1..m} |c_i\rangle \langle c_i|$, $I_c=I_{2^m}$.

In 2WQC we instead perform partial projection on such last $m$ qubits: $\rho \to (I_s\otimes \Pi_c) \rho (I_s\otimes\Pi_c)$, then measure first $n$ qubits:
$$ \textrm{2WQC:}\quad P(s|c)=\frac{\textrm{Tr}((\Pi_s\otimes I_c) (I_s\otimes \Pi_c) \rho (I_s\otimes\Pi_c)}{\sum_s\textrm{Tr}((\Pi_s\otimes I_c) (I_s\otimes \Pi_c) \rho (I_s\otimes\Pi_c))}=$$
\be=\frac{\textrm{Tr}((\Pi_s\otimes \Pi_c)\,\rho)}{P(c)}\qquad\textrm{as in p1WQC}\ee
Hence the final probability distribution is the same as postselected 1WQC, but in 2WQC it is done by projection as in state preparation - physically enforcing the initial/final values.

Therefore, while p1WQC has $P(c)\sim 2^{-m}$ success rate of its postselection, 2WQC should provide the same statistics with success rate close to 1, what can become exponential speedup. For example in the discussed 3-SAT solver: possible with postselection, but requiring exponential numbers of runs.

\subsection{3-SAT problem}
3-SAT~\cite{3SAT} (boolean satisfiability problem) is one of the basic NP complete problems, also very convenient for the discussed approach. We ask if there exist binary variables $x$, such all given alternatives of triples of variables (some can be negated) are satisfied. For example:
$$\exists_{x_1 x_2 \ldots}\ (x_1 \vee\neg x_2 \vee x_3) \wedge (\neg x_4 \vee x_2 \vee \neg x_3) \wedge (x_5 \vee \neg x_4 \vee x_2)\wedge \ldots\ ?$$
For a chosen instance of 3-SAT problem, let us denote $n$ as the number of variables: $x\in\{0,1\}^n$, and $m$ as the number of its alternative clauses - the formula to satisfy is $\forall_{i=1..m}\, C_i(x)$, where $C_i(x)$ clause is alternative of some 3 of variables, some of which can be negated.

Restringing the number of times a variable is used, turns out for 4 appearances the problem is still NP complete~\cite{rSAT}, hence we can assume each variable is used at most 4 times, what allows to bound depth of the proposed 2WQC solver.
\subsection{Basic 2WQC 3-SAT solver}
In Shor algorithm we first prepare superposition of all $2^n$ inputs, then calculate a classical function ($y^a \mod N$ to factorize $N$, using a random $y$) into additional qubits, which are measured - what restricts the ensemble to usually a periodic one. Performing its Fourier transform (QFT) and measuring allows to obtain this period, which provides hint for the factorization problem.

The proposed 3-SAT solver is similar, for the classical function using the verifier of NP problem: for 3-SAT calculating $m$ alternative clauses. We could postselect them to 1 as "true": ensuring satisfaction of all clauses - measuring input qubits we should  get a solution of this 3-SAT problem.

However, such p1WQC approach would have pessimistically $P(c)\sim 2^{-n}$ success probability of postselection, requiring exponential number of runs. 2WQC would allow to replace this $|1\rangle \langle 1|$ postselection with $\langle 1|$ postparation as CPT analog of state preparation, in theory increasing success probability to 1.

Such solver requires 4-qubit C-OR gates (reversible), to be optimized with basic quantum gates in the future.
\be (x,y,z,t)\to \left(x,y,z,t \oplus ((\neg)x \vee (\neg)y \vee (\neg)z)\right) \ee

\subsection{Imperfection model}
For imperfection analysis, it is crucial to understand the number of inputs which satisfy all but $k$ clauses:
\be S(k)=\left\{x\in \{0,1\}^n:| \{i: \neg C_i(x)\}| = k \right\} \ee
where $|\cdot|$ denotes size of a set. Obviously $\sum_{k=0}^m |S(k)|= 2^n$. The solution we search for is in $S(0)$, we can assume there is only one: $S(0)=\{\bar{x}\}$. The remaining $S(k)$ contain false solutions, the lower $k$, the better (but not necessarily useful).

In the proposed basic 2WQC 3-SAT solver, Hadamard gates ideally prepare $\sum_{x\in \{0,1\}^n} |x\rangle /\sqrt{2^n}$ state. Acting with C-ORs and $\langle 1|$ on all $m$ clauses and normalizing, ideally the state should become $|\bar{x}\rangle$, to be measured providing the solution.

However, imperfections of practical realizations will introduce errors. Let us propose a simple \textbf{imperfection model} with $\epsilon$ corresponding to probability of C-OR with $\langle 1|$ allowing incorrect value, suggesting to consider prepared density matrix to be measured as:

\be  \rho \propto \sum_{k=0}^m \epsilon^k \sum_{x\in S(k)} |x\rangle \langle x| \ee

To ensure its measurement gives $\bar{x}\in S(0)$ solution with a high probability, we need to reduce $\epsilon$ to $\sim 2^{-n}$ scale, what rather requires error correction techniques.
\subsection{Error correction techniques to reduce $\epsilon$ to $\sim 2^{-n}$ scale}
To reduce this $\epsilon$, using $k$ copies of C-ORs with $\langle 1|$, all $k$ would need to allow for unsatisfied alternative clause, reducing probability of being incorrect $\epsilon \to \epsilon^k$. Hence using $O(n)$ copies of C-ORs with $\langle 1|$, in theory should allow to reduce $\epsilon$ to a safe error level.

However, such $O(n)$ \textbf{serial copies} would increase depth, hence other issues like decoherence could appear. Alternatively, we can use \textbf{multiple copies of variables in parallel} to reduce depth. The basic quantum error correction~\cite{QEC} approach is copying variables with Controlled-NOT gate:
\be |0\rangle |0\rangle \xrightarrow{H\otimes I} \frac{1}{\sqrt{2}} (|0\rangle+|1\rangle) |0\rangle \xrightarrow{\textrm{C-NOT}} \frac{1}{\sqrt{2}} (|00\rangle+|11\rangle) \ee
In standard quantum error correction such copies would be measured, controlling further processing. In 2WQC we additionally have $\langle 0|$ operation, bringing also new error correction capabilities, for example frequent mid-circuit enforcement of syndrome of error correction code to zero, trying to stabilize the process to use superposition only of codewords - being in a large Hamming distance, hence making errors less likely.

Let us consider \textbf{equalizer gadget} like in Fig. \ref{3SAT} as the simplest tool of this type (enforces syndrome to zero for $\{00,11\}$ codebook). It uses Controlled-XOR gate, which can be realized with 2 Controlled-NOT gates, for which controlled qubit we apply not only state preparation $|0\rangle$, but also its symmetric counterpart $\langle 0 |$. While the former is similar as popular in quantum error correction, usually being measured, in 2WQC we can instead directly project it to $\langle 0 |$, what  seems to bring powerful improvement of error correction capabilities.

Such equalizer should physically enforce equality of two qubits, mathematically projecting them on $(|00\rangle + |11\rangle)/\sqrt{2}$. Using it multiple times should allow to maintain equality through the calculations, e.g. of multiple copies of each variable to be used in copies of C-ORs with $\langle 1|$. Finally, all such copies of variables should be measured, ideally leading to equal values.

Equalizers (EQ) in theory could be also used before other operations as C-ORs, to prepare in 2 layers of equalizers (or more for denser) as many copies of variables as needed: by Hadamards followed by mesh of equalizers:
\be |0\rangle |0\rangle \xrightarrow{H\otimes H} \frac{1}{2} (|0\rangle+|1\rangle) (|0\rangle+|1\rangle) \xrightarrow{\textrm{EQ}} \frac{1}{\sqrt{2}} (|00\rangle+|11\rangle) \ee
The higher the noise, the denser equalizer mesh could be used. Optimization of details, together with better imperfection models, will require further work.

\section{Conclusion and further work}
This article proposes construction of 3-SAT solver for 2WQC, with basic analysis and dedicated error correction mechanisms, in theory allowing to overcome imperfections, and hopefully leading to practical realizations.

There are many direction for further work, for example:
\begin{itemize}
  \item There was used C-OR as 4-qubit gate, which needs reduction and optimization into basic gates.
  \item There was used some basic imperfection model, which should be extended in the future, including e.g. decoherence, and analysis of imperfections of proposed equalizers.
  \item Practical realizations might require optimization of e.g. serial-parallel setting and mesh of equalizers.
  \item While we have focused on 3-SAT, it might be worth considering other NP complete problems, especially having in mind polynomial cost of transformations between such problems.
  \item Finally such NP solvers should be realized and tested: first in simulators, and then in hardware.
  \item We should investigate also possibility of attacks on PSPACE problems, e.g. adding forall quantifier for some variables of 3-SAT problem.
  \item Search for practical applications for such NP solvers, depending on a growing number of available qubits.
  \item Generally, possibility of such quantum NP solvers should be verified, leading to necessary preparations if positive - especially in search for some nextgen postquantum cryptography: resistant to imperfect quantum NP solvers.
\end{itemize}

\bibliographystyle{IEEEtran}
\bibliography{cites}
\end{document}